\documentclass[11pt]{article}
\usepackage{graphicx}
\usepackage{amssymb}
\usepackage{amscd}
\usepackage{mathrsfs}
\usepackage{longtable,lscape}
\usepackage{amsthm}
\usepackage{amsfonts}
\usepackage{amsmath}
\usepackage{bbm}
\usepackage{float}
\usepackage{url}
\usepackage{hyperref}
\usepackage[margin=0.5cm,font=footnotesize]{caption}
\usepackage{lineno}
\usepackage[round]{natbib}
\usepackage{appendix}
\usepackage{subfig}

\begin{document}
\title{\bf From Quantum Cognition to Conceptuality Interpretation II: Unraveling the Quantum Mysteries}
\author{Diederik Aerts$^*$, 
Massimiliano Sassoli de Bianchi\footnote{Center Leo Apostel for Interdisciplinary Studies, 
        Vrije Universiteit Brussel (VUB), Pleinlaan 2,
         1050 Brussels, Belgium; email addresses: diraerts@vub.be,autoricerca@gmail.com} 
        $\,$ and $\,$  Sandro Sozzo\footnote{Department of Humanities and Cultural Heritage (DIUM) and Centre CQSCS, University of Udine, Vicolo Florio 2/b, 33100 Udine, Italy; email address: sandro.sozzo@uniud.it}              }

\date{}

\maketitle
\begin{abstract}
\noindent 
An overview of the conceptuality interpretation of quantum mechanics is presented, along with an explanation of how it sheds light on key quantum and relativistic phenomena.
In particular, we show how the interpretation clarifies Heisenberg's uncertainty principle, wave function-based and entanglement-based nonlocality, interference effects resulting from the superposition principle, delayed choice experiments, quantum measurements, the mechanism of quantization, the reason why entities can establish entanglement bonds, and the statistical behavior of indistinguishable entities. We further argue that the interpretation can also elucidate relativistic effects, focusing on time dilation. Finally, we suggest that it can provide a novel and challenging perspective on evolution. This article is the second in a two-part series devoted to exploring this promising approach to reality. The first part, which serves as a companion to this discussion, outlines the intellectual trajectory leading from the first applications of quantum notions to human cognition to the bold rethinking suggested by the conceptuality interpretation.
\end{abstract}

\section{Introduction\label{intro}}

Quantum mechanics remains profoundly difficult to interpret, often regarded as the ``most enigmatic theory,'' due to its departure from classical space-time conception of the physical world. Quantum entities do not appear to possess, in actual terms, properties ensuring a stable presence in space and time, e.g., a jointly well-defined position and momentum. Moreover, the standard quantum formalism lacks a unanimously accepted interpretation, i.e., a conceptual framework to explain the counterintuitive behavior of microscopic entities. The \emph{conceptuality interpretation of quantum mechanics}, proposed by one of us in 2009, attempts to fill this explanatory gap by introducing the speculative hypothesis that quantum entities are \emph{carriers of meaning}, so that their interaction with each other and with measuring apparatuses can only become intelligible if we accept that all these interactions are \emph{meaning driven}, similar to what happens when humans communicate through natural language \citep{aerts2009,Aerts2010a,Aerts2010b,Aerts2013,Aerts2014,aertssassoli2018,aertssassolisozzoveloz2019,aertsetal2020,aertsbeltran2020,sassoli2021,aertssassoli2022,aertssassoli2024b,aertssassolisozzo2024}. 

To understand the path that led to the formulation of the conceptuality interpretation, which we have outlined in detail in the first part of this two-part article \citep{aertssassolisozzo2024}, one has to start from the successes of the \emph{quantum cognition} program. Among the authors who contributed most to its development, we can mention, in addition to the members of our Brussels group (see all the names mentioned in \citet{aertssassolisozzo2024}), Andrei Khrennikov and Harald Atmanspacher for its first developments, then Jerome Busemeyer, Reinhard Blutner, Peter Bruza, Emmanuel Haven, and Emmanuel Pothos for its expansion into a full-fledged research field, to name but a few; see \citet{khrennikov2010,busemeyerbruza2012,pothos2013,haven2013,ashtiani2015,bruza2015,wendt2015,pothos2022} and the references cited therein. 

In a nutshell, this program attempted to use the conceptual and mathematical framework of quantum mechanics to model human cognition, e.g., decision-making processes, and how meaning emerges from conceptual combinations. The scale of this success was in some ways unexpected, so much so that at a certain point the idea of a reversal presented itself. The situation was reminiscent of that of Louis de Broglie at the beginning of the last century, who was confronted with the observation that entities usually thought to be wave-like could be described with advantage, in certain experimental contexts, by assuming that they also possessed a corpuscular nature, as proposed by Planck and Einstein. From this observation, a speculative idea emerged in the French scientist's mind: that things could also function in a specular manner, i.e., entities usually thought to possess a corpuscular nature, such as electrons, could also manifest wave-like behavior \citep{debroglie1924}. This vision, which initially met considerable resistance, had an extraordinary impact on quantum mechanics, paving the way for the wave interpretation of matter and contributing, a few years later, to Schr\"{o}dinger wave mechanics.

The conceptuality interpretation can be seen as an example of a ``move \`a la de Broglie.'' In fact, the success of the quantum cognition program was telling us that humans think and make decisions pretty much in a quantum-like way. Of course, this doesn't mean that quantum processes necessarily take place at the level of the brain, but rather that our thought processes, as manifested with the help of our brain activity, are organized according to quantum structures. Now, if human conceptual entities possess a quantum-like behavior, and thus have, in a sense, a quantum nature, one may be temped to image the reverse, that microscopic quantum entities possess a conceptual-like behavior, hence, in a sense, a conceptual nature similar to that of human concepts. Unlike wave-particle duality, however, the relationship between \emph{quantumness} and \emph{conceptuality} would not be complementary, but rather based on similarity, as two notions describing the same aspect, which can reveal itself at various organizational levels.

However, it was not only the success of quantum cognition and a reversal \`a la de Broglie that gave rise to the idea of the conceptuality interpretation, and two other factors should be mentioned. Firstly, the growing number of experimental results suggesting that quantum entities are not always located in space and time, and an important example was the neutron interferometer experiments performed by Rauch \emph{et al} \citep{Rauchetal1975} and Werner \emph{et al} \citep{Werneretal1975} in the 1970s, which succeeded in delocalizing a neutron over several centimeters. Secondly, there was the realization that a concept is also an entity characterized by its properties, as put forward in the 1970s by Eleanor Rosch in her \emph{prototype theory} \citep{rosch1973}). Thus, it was additionally suggested that the observed non-spatiality of quantum entities could be due to their conceptual nature. 

It is also important to note that the conceptuality interpretation is not aligned with idealistic philosophy, where physical theories are seen as merely theories of human mental content. Instead, it adopts a \emph{realist view} that treats conceptual entities as real entities that exist in different layers of our reality, which can undergo measurement processes, the latter being meaning-driven interactions through which one can discover pre-existing properties but also create new ones. This observation helps clarify that the conceptuality interpretation, while giving importance to the representation and communication of \emph{information}, is not an information-theoretic interpretation \citep{wheeler1990,caves2002,feldmann2023} but an approach suggesting that the nature of reality is intrinsically conceptual, hence cognitive. In other words, while its realism is rooted in operationality, its focus lies on the ontological aspects of the world. This implies that insights from human cognition could help decipher the behavior of quantum (and relativistic) entities, provided we carefully distinguish between quantum entities, studied by experimental physicists, and human conceptual entities, studied by cognitive scientists.

That being said, the scope of this article is to  explore, one by one, various quantum (and relativistic) phenomena, each time showing how they are addressed in a much more natural and intelligible way by adopting a conceptualistic-cognitivistic approach, that is, on the assumption that quantum entities are \emph{conceptual} rather than \emph{objectual}. More specifically, in Section~\ref{uncertainty}, we discuss Heisenberg's uncertainty principle and nonlocality. In Section~\ref{quantumsuperposition}, we consider the interference effects produced by quantum superposition and delayed choice experiments. In Section~\ref{quantummeasurement}, we consider quantum measurement and the phenomenon of quantization, and in Section~\ref{quantumentanglement}, quantum entanglement and the statistics of indiscernible entities. Finally, in Section~\ref{timedilation}, we explain how time dilation also follows from a conceptualist approach  \citep{aerts2018,aertssassoli2024a}, and in Section~\ref{conclusion}, we offer a more speculative perspective on how the latter might radically change the way we understand reality.

\section{Uncertainty and nonlocality\label{uncertainty}}

Heisenberg's uncertainty principle suggests that the values of certain (complementary) physical quantities (observables), such as position and momentum, when they refer to quantum entities, cannot both be predicted with certainty, not because we lack knowledge of their actual values, but because they do not have actual values before they are measured. More precisely, if we measure one quantity, such as position, and actualize a particular value for it, we exclude the possibility of doing the same, simultaneously, for another quantity, such as momentum. 

That incompatibility between position and momentum occurs, not at the level of knowledge but, rather, at the level of definiteness, of their values, is evident in the quantum formalism, as there are no common eigenstates of these two non-commuting observables, whereas this is always the case for a classical entity. One of the mysteries of quantum physics is therefore to understand why this is the case. To emphasise this ontological aspect of Heisenberg's uncertainty relations, it is important to note something less known, namely, a reverse form of the uncertainty relations also holds, i.e., there are no states for a quantum entity such that the values of two complementary observables, such as position and momentum, become jointly highly unpredictable, in the sense that, as the value of one becomes more predictable, the predictability of the value of the other correspondingly decreases, and vice versa \citep{mondal2017}. So why is it that micro-entities such as photons, electrons, quarks, etc. cannot exist in states where complementary observables are simultaneously maximally sharp or, conversely, maximally unsharp?

The conceptuality interpretation provides a straightforward answer: if micro-entities are conceptual in nature, they cannot occupy states that are simultaneously maximally \emph{abstract} and maximally \emph{concrete}. A maximally concrete state implies a maximal spatial localization, while a maximally abstract state implies a maximal despatialisation. Thus, Heisenberg's uncertainty principle would be inherently consistent with the conceptuality interpretation, and therefore ontological, and the same holds for the reverse form of Heisenberg's relations, since it is obvious that, if micro-entities are conceptual, they also cannot exist in states that are minimally abstract and minimally concrete at the same time.

Let us illustrate what we mean with an example. Consider the abstract concept \emph{Horse}, on the one hand, and the conceptual combination \emph{The Brabant draught horse named Iltschi of the Diepensteyn Stud Farm in Belgium}, on the other hand. The former is a very abstract concept, delocalized with respect to all actual material horses, and its quantum representation would be that of a plane wave, with a well-defined momentum and a totally indeterminate position. The latter is instead a very concrete concept, corresponding to a unique material horse, and its quantum representation would be that of a delta function.\footnote{In this example, for didactic purposes, we have compared the conceptual entities of human language with the entities of the material world, which, according to the conceptuality interpretation, are themselves conceptual in nature, but should not be confused with the former. See also the discussion in \citet{aertssassolisozzo2024}.}

The fact that a quantum entity, in general, does not have a definite position in space (or a definite momentum) until it is measured, is an expression of \emph{nonlocality}. While often exemplified through the phenomenon of entanglement, where measurements on one entity cannot be separated from measurements on a second entity, even if the two are separated by large distances (see Section~\ref{quantumentanglement}), the notion of quantum nonlocality extends to the entire structure of quantum mechanics, as a consequence of the superposition principle but also of the very interpretation of the wave-function as describing the probability amplitudes of the presence of a quantum entity in different spatial locations, which generally spreads extremely fast over time. Indeed, the information encoded in the wave function is nonlocal in the sense that it expresses the potentiality of a quantum entity, at a given time, to lend itself to the creation of different positions.  

The above is incompatible with the classical idea that a physical entity should locally exist in space and time, occupying a specific location in space at each instant in time. In other words, not only experimentally, as with Rauch's neutron interferometer experiments, but also theoretically,  quantum mechanics tells us that quantum entities can be in states of unactualized spatial properties, i.e., in \emph{non-spatial} states. Understanding the origin of non-spatiality, which gives rise to the phenomenon of nonlocality, is therefore another of the quantum mysteries. What does it mean to be in a non-spatial state? The answer provided by the conceptuality interpretation is that non-spatiality is just an expression of the conceptual (abstract) nature of physical entities.  

For example, where are human languages located? We do not refer here of the traces left by a language, that we can find for instance in books and other memory supports. There is indeed also an \emph{abstract} aspect of a language, understood as a pure \emph{conceptual} entity, not reducible to its possible spatializations, and that aspect is what we usually call \emph{meaning}. If a quantum entity is understood as a conceptual entity that carries meaning, which is something very different from a printed word (or combination of words) that is to be regarded only as the trace left by such a meaning, it becomes clear, and even self-evident, why it cannot be spatial in nature. Consider the distinction between the statements: ``At this moment, Massimiliano is visiting Diederik in Brussels,'' and ``At this moment, Massimiliano is visiting either Diederik in Brussels or Sandro in Udine.'' The first sentence describes a localized state of the conceptual entity Massimiliano (distinct from Massimiliano's physical presence), whereas the second describes a delocalized superposition state of Massimiliano's potential presence in two distinct places. In the conceptual realm, the latter poses no mystery. If quantum entities are fundamentally conceptual, neither their non-spatial nature nor the associated superposition states should then be seen as puzzling.

\section{Interference and delayed choice\label{quantumsuperposition}}

Another quantum mystery concerns the interference effects produced by superposition states. The most famous illustration of this phenomenon is provided by the double-slit experiments \citep{thomas1804, taylor1909}, which have been repeated over the decades not only with photons, but also with electrons, neutrons, atoms and even large molecules. In these experiments, quantum entities appear to interfere with themselves, as if they could pass through both slits at once. It is certainly possible to attempt to explain the observed interference figures by assuming that quantum entities are physical waves capable of jointly crossing both slits. However, such an explanation proves insufficient when one observes that the traces formed on the detection screen, over time, are extremely localized impacts, as if left by entities of a corpuscular nature. 

According to the conceptuality interpretation, the wave aspect associated with the entities passing through the slits, mathematically described by a wave function evolving according to the Schr\"{o}dinger equation, is just a convenient way to model, by means of constructive and destructive interference effects, the different overextension and underextension probabilistic effects resulting from the cognitive processes that take place in the course of the experiment. We can indeed think of the detection screen as the cognitive entity that, upon the arrival of each photon, electron, neutron, etc., must answer a specific question, providing its answer by means of a localized impact. Translated into human language, the question would essentially be the following: ``What is a good example of an impact point of an entity passing through one of the two slits?''. As we will explain, by answering such a question and staying on a purely conceptual (abstract) level, it becomes possible to understand the fringe pattern that emerges. 

Let us focus first on the central fringe, which is equidistant from the two slits and is where the impact density is the highest. This area clearly best exemplifies the concept of ``an entity passing through one of the slits,'' because it reflects a maximum uncertainty about which slit was used. On the other hand,  in the two regions opposite the slits, the impact density is the lowest, as these regions minimize the uncertainty about the entity's path, making them poor examples of the concept in question. Moving then outward from the center, we are back again in a situation of uncertainty, although less important than that expressed by the central region, so there will be new fringes, but with a lower density of impacts. This explanation gains more credibility when we consider that (i)  even in the human cognitive domain it is possible to consider situations similar to the double-slit, and (ii) such situations are also capable of producing interference effects, although wave-like phenomena are clearly not present. Let us briefly describe an experiment where this has been explicitly demonstrated, referring the interested reader to \citet{aerts2009} for the details.

\begin{figure}
\begin{center}
\includegraphics[scale=0.37]{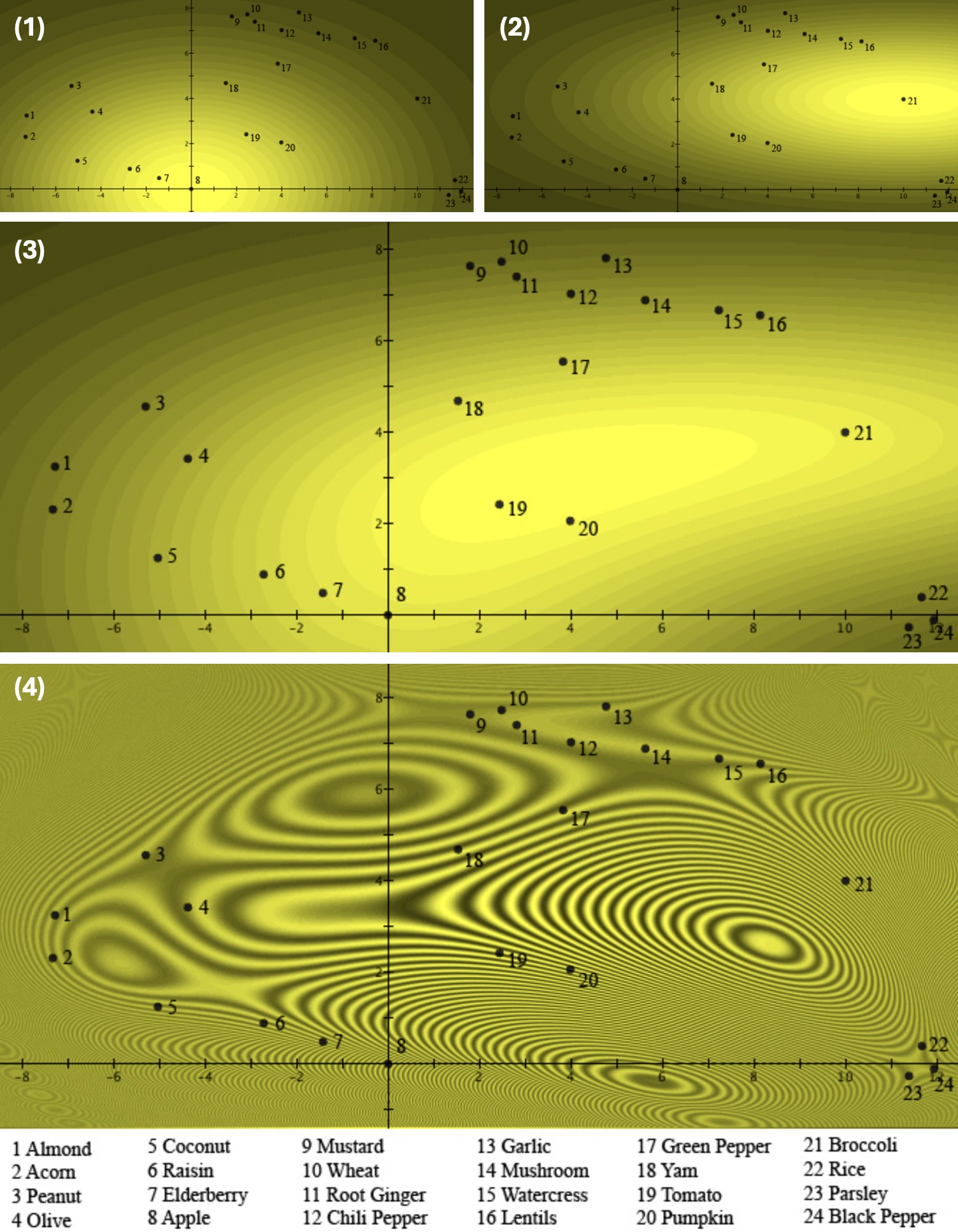}
\end{center}
\vspace{-15pt}
\caption{Figure (1) represents the probabilities of selecting a good exemplar of  \emph{Fruit}, with \emph{Apple} (number 8) being the most frequently chosen. Figure (2) represents the probabilities of selecting a good exemplar  of  \emph{Vegetable}, with \emph{Broccoli} (number 21) being the most frequently chosen. Figure (3) represents the probabilities obtained by uniformly averaging the probabilities of figures (1) and (2). Figure (4) represents the probabilities of selecting a good exemplar of  \emph{Fruit or vegetable}, which are  markedly different from those of figure (3), revealing the overextension and underextension effects in Hampton's data. For more details on how these figures were obtained, see \citet{aerts2009}.}
\label{Figure1}
\end{figure}

In the late 1980s, the psychologist James Hampton performed a series of cognitive test. In one of these, he presented a list of $24$ exemplars to $40$ students, asking them to judge, for each exemplar, whether they considered the exemplar as a member of (a) \emph{Fruit}, (b) \emph{Vegetable}, (c) \emph{Fruit or Vegetable} \citep{hampton1988}. In this setup, the various exemplars of \emph{Food} work similarly to the different positions on a detection screen in the double-slit experiment, with the concepts \emph{Fruit} and \emph{Vegetable} corresponding to the two slits. Now, if the students' decision-making for question (c) had followed a sequential process -- first choosing between \emph{Fruit} and \emph{Vegetable} and then, based on this choice, selecting a representative exemplar of either \emph{Fruit} or \emph{Vegetable} -- then the probability of selecting any given exemplar of \emph{Food} would have aligned with the average of the probabilities from questions (a) and (b). However, Hampton's data did not support this. Instead, it displayed a complex mix of overextension and underextension effects, i.e., of values above and below the simple average of situations (a) and (b), respectively. When Hampton's data were suitably analyzed and represented in a quantum mechanical model, using two $2$-dimensional complex wave functions to model the responses to questions (a) and (b), and then combining these functions in a normalized superposition to represent question (c), an interference pattern emerged, similar to those observed in birefringence phenomena \citep{aerts2009} (see Figure~\ref{Figure1}).

It is important to note that the interference pattern in Figure~\ref{Figure1} is generated by each subject in Hampton's experiment by choosing a particular exemplar based on the meaning of the \emph{Fruit or vegetable} conceptual combination and the way it is understood. It is therefore a process that takes place at an abstract, conceptual level, within each subject's mental structure. Thus, each of them, with her or his response, is able to statistically bring the entire interference figure to life. Of course, they all have their own way of choosing an exemplar, but the average among all their \emph{ways of choosing} will tend, probabilistically speaking, toward the \emph{Born's rule} prediction. Such an average has been called \emph{universal average} and it can be shown that, if the state space is Hilbertian, it does indeed correspond to the quantum Born rule \citep{AertsSassoli2015a,AertsSassoli2014}. Therefore, the latter would only be a first-order approximation of a more general probabilistic rule. Now, the conceptuality interpretation tells us that the construction of the interference figure in a double-slit experiment in the physics lab would happen in exactly the same way, and this would explain how individual impacts on the detection screen are able to generate it, one by one. Each of these impacts would in fact result from the measuring instrument's understanding of the  meaning expressed by the quantum entity's conceptual interaction with the experimental setting, as explained above.

As a consequence, the conceptuality interpretation also allows to easily explain the so-called \emph{delayed-choice experiments} \citep{Wheeler1978, Jacquesetal2007}, where one can change the experimental setup at the last moment, choosing either a ``wave setup,'' like the one used in the double-slit experiment, or a ``particle setup,'' compatible with a corpuscular description. Experiments of this kind further demonstrate the inadequacy of the wave-particle duality, but remain compatible with the description of quantum entities as conceptual entities, whose states can change from a more abstract (non-spatial) to a more concrete (spatial) one, when they are drawn into space at the moment of their detection. This means that the answer that the cognitive-like detection instrument will provide, for example in the form of an impact, will simply correspond to the meaning conveyed by the final setup (the different setup corresponding to the different questions). Indeed, the ``wave setup'' is then the equivalent of a ``particle setup'' where the information about the path taken by the entity under consideration has been \emph{erased}. Erasing information means creating a situation of lack of knowledge, and the latter is to be associated with a genuinely new element of reality in the conceptual realm, considering that it is a cognitive entity that performs the measurement. The process of information erasure is therefore equivalent to an objective change of the state of the conceptual entity, which becomes a superposition state, and this explains the appearance of the interference effects.

\section{Measurement and quantization\label{quantummeasurement}}

Let us consider the measurement of an observable performed on a quantum entity prepared in a given state. If the latter is a superposition state with respect to the measured observable, the measurement will in general produce a \emph{collapse} of the state into one of the eigenstates of the measured observable. The collapse appears both irreversible and unpredictable, challenging causal and deterministic views of the physical world. Explaining such a process as a quantum physical process is what has been called the \emph{measurement problem}, which remains one of the quantum mysteries awaiting elucidation. As we have partly explained, the conceptuality interpretation, consistently with how we understand human decision-making, views measurements as cognitive processes, where the apparatus is considered a structure sensitive to the meaning carried by the measured quantum entity, understood as a conceptual entity, and subjected to an interrogative context where it is required to provide an answer. However, this is not just a clever analogy. Indeed, a human decision process can be understood as a \emph{tension-reduction} process where the actual (the decision taken) breaks the symmetry of the potential (the available options), and such tension-reduction -- weighted symmetry breaking -- process can be found ``hidden'' in the quantum formalism, when reformulated using Bloch's extended representation. Quoting from \citet{AertsSassoli2015a}:

{\quote [...] when a human subject is confronted with a question [...], and an associated set of $N$ possible answers, this will automatically build a mental (neural) state of equilibrium, which results from the balancing of the different tensions between the initial state of the concept subjected to the question, and the available mutually excluding answers that compete with each other. [...] at some moment this mental equilibrium will be disturbed, in a non-predictable way, and the disturbance will cause an irreversible process during which, very quickly, the initial conceptual state will be drawn to one of the possible answers. [...] This \emph{tension-reduction process}, however, will not always result in a full resolution of the conflict between all the competing answers. There are contexts such that the state of the system can be brought into another state of equilibrium, between a reduced set of possibilities. \endquote}

The second situation described in the above passage is that characteristic of so-called \emph{degenerate measurements}, where the measured entity remains in a state of superposition within a given subspace of the Hilbert space. Now, the above description of what happens, cognitively speaking, in the course of a human decision process, exactly mirrors the description of a quantum measurement according to the \emph{hidden-measurement interpretation}, which uses the mathematical language of the \emph{extended Bloch representation} (EBR) of quantum mechanics \citep{AertsSassoli2014}. 

More precisely, the EBR is a way to reformulate the standard quantum formalism using a generalization and extension of the historical three-dimensional Bloch sphere model. In this representation, it becomes possible to associate real unit vectors with the initial state of the measured entity, and with the available outcome states. These real vectors are $(N^2-1)$-dimensional if the associated Hilbert space is $N$-dimensional. The vectors representing the outcome states are then the vertices of an $(N-1)$-dimensional simplex $\triangle_{N-1}$, inscribed in the convex region of the states, which in turn is inscribed in an $(N^2-1)$-dimensional unit sphere. In this setting, it can be shown that there is a first phase in a quantum measurement which corresponds to the immersion of the initial vector state deep inside the convex region of states, along a path orthogonal to $\triangle_{N-1}$, until it reaches an on-simplex equilibrium point ${\bf r}_e\in \triangle_{N-1}$. This corresponds to the construction of the previously mentioned mental equilibrium state, where the conceptual entity is brought into contact with the ``potentiality region'' generated by the $N$ mutually exclusive answers. This results in a partitioning of $\triangle_{N-1}$ into $N$ convex subregions, formalizing the unstable tensional equilibrium. Imagining these regions to be filled with an abstract elastic and disintegrable substance, then an unpredictable perturbation (fluctuation) will cause one of them to collapse, with the consequence that the equilibrium state ${\bf r}_e$ will be brought towards the corresponding vertex vector, thus producing the outcome. 

In the limited space of this article we cannot describe this process in detail, which also generalizes to the case of degenerate measurements.  What however is important to emphasize is that, if one calculates the probabilities of the different tension-reduction processes, via the  disintegration of the corresponding subregions, one recovers exactly the \emph{Born rule} \citep{AertsSassoli2014,AertsSassoli2015a}. Thus, the derivation of the quantum probabilistic rule as a result of an interaction mechanism between the measured entity and the measuring apparatus is perfectly compatible with our human intuition of what happens at the intrapsychic level during a decision-making process, which provides an additional argument in favor of the conceptuality interpretation. 

It is important to emphasize that the ideas we have just discussed can be conveyed through the precise and powerful mathematical framework of the EBR. Note also that the hidden-measurement solution to the measurement problem  was originally developed from perspectives unrelated to the conceptuality interpretation. However, in retrospect, we can see that the interpretation inherently encompasses all the essential components necessary to lead to such interpretation. Of course, the transition from the intrapsychic description of a decision-making process, as we humans experience it, to its mathematical formalization within the EBR framework, is not at all evident, and required being able to identify specific mathematical structures within a Hilbert space, and we refer the reader to  \citet{AertsSassoli2014} for the details. 

What we additionally want to emphasize is that the above framework contains the possibility of describing another important cognitive phenomenon, namely, 
 \emph{categorical perception} \citep{goldstone2010}, which can be understood as the cognitive equivalent of the phenomenon of \emph{quantization}, to be here understood in the sense historically attributed to it by Max Planck, in the article that initiated Old Quantum Theory \citep{planck1900}. Note that Planck's quantization was originally a theoretical device, without a clear physical interpretation, but even to this day it remains unclear why quantization takes place in physical reality. Of course, starting from the wave description via the Schr\"{o}dinger equation, it is easy to derive the discrete spectrum of the Hamiltonian as a consequence of the confinement of a quantum entity, like in the simple example of a box. Indeed, when a quantum entity is confined, its wave function must satisfy specific boundary conditions, which in turn impose limits on the possible wavelengths, and therefore on the permitted energy states. But according to the conceptuality interpretation, the wave function is just a convenient mathematical  tool for modeling the underlying cognitive reality. 

We can think of the confining box as a cognitive entity that \emph{perceives} the existence of the conceptual entities that evolve inside it. The emergence of a discrete spectrum would then be similar to the cognitive mechanism that makes us see discrete colors in a continuum of electromagnetic frequencies. This systematic warping of stimuli by human perception, called categorical perception, occurs in all forms of human perception between what psychologists call \emph{stimuli}, on the one hand, and what they call \emph{percepts}, on the other. This hypothesis is strengthened by observing that the seed of the categorical perception mechanism can also be found in a quantum measurement, in the sense that it would be the decoherence and collapse of a pre-measurement state that would produce it \citep{aertsaerts2022,aertssassoli2024b}. 

Without going into details here, let us simply state that the natural notion of distance between states, when they are on the surface of the convex region containing them in the EBR, where the stimuli lie, before they collapse onto the simplex describing a particular measurement, is different from the natural notion of distance between the same states when they immerse and reach their equilibrium point on the simplex, where the percepts lie. This difference in distances, which means that pairs of stimuli separated by a same distance can be associated with pairs of percepts whose distances may be closer or farther apart, depending on their location, would be precisely the quantum equivalent of categorical perception, which in turn could be at the origin of the phenomenon of quantization. In other words, categorical perception would be built into the very geometry of a quantum measurement.

\section{Entanglement and indistinguishability\label{quantumentanglement}}

Let us now move our attention to the phenomenon of entanglement, where two quantum entities behave as a single interconnected whole, which can also be easily understood in the conceptuality interpretation. Indeed, the perceived ``spookiness'' of the entanglement phenomenon would not be that of an unexplainable ``action at a distance,'' but the result of conceptual entities being necessarily \emph{connected through meaning}. 

To understand why this is sufficient to explain how quantum correlations are created, in typical Bell-type experiments, it is sufficient to observe that when two objects, say two dice, are connected by a third entity, say a rigid rod glued on two of their opposing faces, then when jointly acting on them, for example rolling them, correlated outcomes will be obtained, and it is easy to show that this is sufficient to violate Bell's inequalities \citep{sassoli2013,aabgssv2019}. So, to explain entanglement one needs to explain what would be the element of reality playing the same role played by a rigid rod connecting two dice. Such element, however, cannot be spatially manifest, as we know that nothing can be found in between two entangled entities that are able to fly apart for very large distances, still remaining mysteriously interconnected. The conceptuality interpretation tells us that this non-spatial connection is a \emph{meaning connection}, and indeed, when analyzing an entangled composite system using the EBR, the formalism makes it possible to clarify not only that the two entangled sub-entities are in well-determined states (represented by density operators), but also that it is possible to associate the equivalent of a state with their mutual connection, described in the EBR by a vector of higher dimensionality, expression of the fact that their connection would correspond to a more abstract element of reality \citep{AertsSassoli2016}. 

The above becomes easy to understand when we use the example of human conceptual entities and the way they can connect in terms of meaning. Consider the following short text: \emph{In the zoo, the animals act in different ways, filling spaces with echoes. Children listen, faces bright, to the murmurs and voices and rhythms. The zoo hums, as life pulses around, alive with sounds and motions. Children stand still, absorbed by sounds from every direction, feeling the presence of the animals}. This text can be considered to be a context for the two concepts \emph{Animal} and \emph{Acts}, which in our human culture are clearly meaning connected. Indeed, we all know, from our experience of the world, that there are actions certain animals will typically do that other animals will not do. This meaning connection can be analyzed quantitatively when specific couples of exemplars of these two  concepts  are considered, like the following: $A$ = (\emph{Horse}, \emph{Bear}), $A'$ = (\emph{Tiger}, \emph{Cat}), for \emph{Animal}, and $B$ = (\emph{Growls}, \emph{Whinnies}), $B'$ = (\emph{Snorts}, \emph{Meows}), for \emph{Acts} \citep{as2011}. If we then ask individuals to select pairs of these exemplars for \emph{Animal} and  \emph{Acts}, considered to be representative of their combination as per the above text, when choosing from the couples $A$ and $B$ their selection will be the outcome of a joint measurement $AB$, of the two conceptual entities \emph{Animal} and \emph{Acts}, and similarly for the other possible combinations, defining in total four joint measurements: $AB$, $A'B$, $AB'$ and $A'B'$. The statistics of the outcomes of these joint measurements can then be analyzed using the Bell-CHSH  inequality\footnote{The Bell-CHSH inequality 
can be expressed as $E(A',B')+E(A',B)+E(A,B')-E(A,B) \le 2$, where $E(A,B)$ denotes the expectation value for the joint measurement $AB$, given by $E(A,B)=p(A_1,B_1)-p(A_1,B_2)-p(A_2,B_1)+p(A_2,B_2)$, with $p(A_1,B_1)$ the probability for obtaining the outcomes $(A_1, B_1)$, i.e., (\emph{Horse}, \emph{Growls}), and similarly for the other probabilities and joint measurements.}   \citep{clauser1969} and the result is that the latter is violated to an extent similar to physics situations with entangled spins or entangled photons \citep{as2011,as2014}.

Entanglement is of course widespread in quantum mechanics, and one area in which it plays a special role is when one deals with \emph{identical} entities. Indeed, identical entities are considered as \emph{physically indistinguishable} in the standard formulation of quantum mechanics, 
as 
expounded in modern manuals of the theory. As a consequence of complete indistinguishability, the states of identical entities should have a defined exchange symmetry, namely, they should be either symmetric or anti-symmetric with respect to the exchange of two entities. It is well known that the symmetric case occurs for entities with integer spin (bosons, e.g., photons), while the anti-symmetric case occurs for entities with semi-integer spin (fermions, e.g., electrons). This mathematical requirement forces the possible states of identical entities to be entangled states, which is typically maintained to be at the basis of the statistical behavior
of identical quantum entities:  \emph{Bose--Einstein statistics} for bosons and \emph{Fermi-Dirac statistics} for fermions. 
More specifically, 
bosons tend to cluster together in the same energy (micro-)states
reaching in the extreme case the 
form of a \emph{Bose-Einstein condensate}, where all entities are in the same (micro-)state. In particular, the distribution of bosonic entities across 
the 
available energy levels is deeply different from the statistical distribution of classical indistinguishable entities, which instead obey the \emph{Maxwell-Boltzmann statistics}. If the conceptuality interpretation is correct, this difference in the way the energy levels are occupied must be due to the bosonic entities being connected by meaning, like the words composing a story. 

Take again the short text above. Words do not appear in it with the same frequencies. For example, one can verify that the word \emph{The} appears $6$ times and is the one that appears the most (rank $1$). The second word appearing the most (rank $2$) is \emph{And}, being present in the text $3$ times. Then we have words like \emph{Zoo}, appearing $2$ times, and there are also words appearing just $1$ time. If we multiply a word's rank by the number of times it appears, for \emph{The} we find $1\times 6=6$, and for \emph{And} we find $2\times 3=6$. If we assign rank $3$ to \emph{Zoo}, we also find  $3\times 2=6$. In other words, the product of the rank and the number of occurrences of a word seems to be equal to a constant. In our example, since the text is very short, this does not work beyond the first three words, but it is actually an empirical law, the so-called \emph{Zipf's law} \citep{zipf1935,zipf1949}.

\begin{figure}
\begin{center}
\includegraphics[scale=0.37]{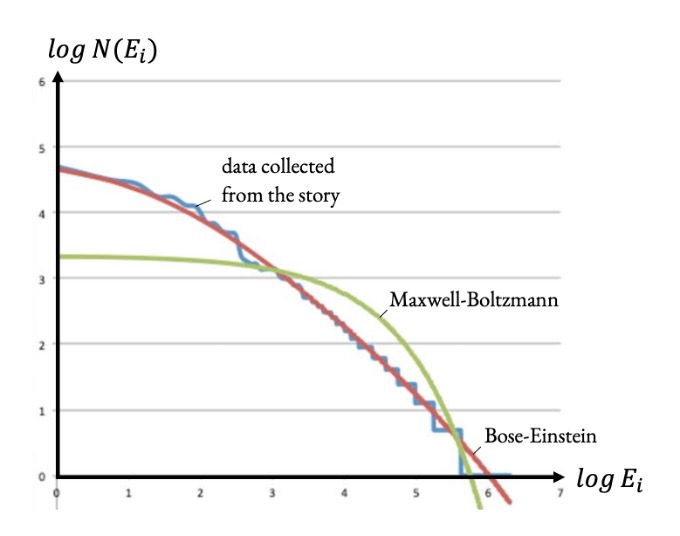}
\end{center}
\vspace{-15pt}
\caption{The log of the numbers $N(E_i)$, of word appearances, in the  Winnie the Pooh  story \emph{In which Piglet meets a Heffalump} \citep{milne1926}, as a function of the log of the associated energy levels $E_i$, compared to the Maxwell-Boltzmann and Bose-Einstein distributions. Only the latter coincides almost completely with the graph of the data. The Figure is adapted from Figure 1(b) of \citet{aertsbeltran2020}.}
\label{Figure2}
\end{figure}

The origin of Zipf's law has always remained an open question. Why does it appear so frequently in so many different areas of human activity, and especially in relation to the frequency with which different words are used in texts that convey a well-defined meaning? The answer provided by the conceptuality interpretation is surprising. If it is true that quantum entities are conceptual, then the relative frequencies with which words appear in texts produced by human language, according to Zipf's law, should be the same as the way in which bosons occupy energy states in a (sufficiently cold and dense) confined gas, according to the Bose-Einstein distribution. In other words, Zipf's law and the Bose-Einstein distribution should be one and the same thing. And in confirmation of this, it is indeed possible to derive Zipf's law, and its generalizations, from the Bose-Einstein distribution. To do this, one simply has to consider the words in a text as  \emph{cognitons}, i.e., as the equivalent of the bosons of a confined quantum gas. The most frequently repeated word will then be associated with the lower energy level of this ``gas of cognitons,'' which will also be the most densely populated. The second most repeated word will populate the first excited level, and so on.  One can then ask what would be a good function for modeling these observed occupancy numbers, for these different energy levels, and the answer is that the perfect function for doing so is the Bose-Einstein distribution \citep{aertsbeltran2020}; see Figure~\ref{Figure2}. 

This connection between Zipf's law and the Bose-Einstein distribution would remain mysterious without the interpretative context provided by the conceptuality interpretation. If we assume that \emph{meaning} is an \emph{abstract substance} that can leave \emph{concrete traces} in our spatiotemporal theater, it will do so through the signature of specific structures, and Zipf's law would be one of them. Thus, we must think of an ideal gas of bosons as the equivalent of a story written with conceptual entities following a distribution that is the same as that followed by the cognitons forming a human story. And the fact that Bose-Einstein statistics is just the equivalent in the physical world of Zipf's law, describing many of the artifacts of human culture, offers a surprising confirmation of this fundamental bridge between linguistic and quantum structures. 

In fact, when a human story is viewed as a collection of cognitons, the phenomena of quantum indistinguishability and quantum entanglement become almost self-evident, and this, in turn, through the conceptuality interpretation, allows to understand the nature of these phenomena in the physical world. Take again the example of the previous text, about the animals in the zoo behaving in different ways. In it, the concept \emph{Zoo} appears twice, in the first and third sentence. It is clear that if we exchange these two concepts in the story, its meaning will not change, not even in the slightest. This invariance is, of course, due to the absolute indistinguishability of the two \emph{Zoo} concepts in the story, and if the latter were represented in a quantum mathematical language, they would correspond to a wave function that is symmetric with respect to the exchange of these two \emph{Zoo} components. 

Also, if we deterministically contextualize the first \emph{Zoo} in the story, combining it with the concept \emph{London}, so that it becomes  \emph{London Zoo}, then also the second \emph{Zoo} will change its meaning, and will do so instantaneously. Of course, if the text is written on paper, and we replace the word ``zoo'' in the first sentence with the word ``London zoo,'' the word ``zoo'' appearing in the third sentence will not change accordingly. Written words are the traces left by conceptual entities and should not be confused with the abstract form of them, the concepts (see for example the discussion in \citet{aertsetalweb2018}). But at the conceptual level, when we change the first \emph{Zoo} into  \emph{London Zoo}, also the second \emph{Zoo} becomes the  \emph{London Zoo}. There is no ``spooky action at a distance,'' as the process is in its essence non-spatiotemporal, i.e., it happens at the abstract level of the meaning structure of the story. And of course, we can also choose to change the second \emph{Zoo} to \emph{London Zoo}, and it will then be the first \emph{Zoo} that will instantaneously change in meaning, as happens in delayed choice experiments. 

We finally stress that, in the conceptuality interpretation, \emph{understanding} is a process of meaningfully combining concepts, similar to how atoms in a Bose gas form a Bose-Einstein condensate. Just as sentences and paragraphs in a text are intelligently combined to create meaning, then leading to understanding, atoms in a Bose gas create superposition states, as their de Broglie waves overlap at low temperatures. Understanding would thus be a \emph{state of integration of meanings}, akin to how atoms, condensing into a unified state in a Bose-Einstein condensate, are able to form new entities, i.e., new \emph{coherence domains}, from the collective superposition. In this respect, the formation of entanglement, both in the physical and in the cognitive-linguistic domain, can be attributed to a mechanism of \emph{contextual updating} \citep{ijtp2023,philtransa2023}. Each time one adds a new word to a text, as in the example above, the new word depends on the meaning of all the other words and contextually updates the meaning content of the entire text. This is because the new word not only has to fit into the overall narrative of the text, but also, by being added, gives new meaning to it. 

The contextual updating mechanism mirrors the \emph{tensor product} procedure used to describe composite quantum entities. Indeed, when a quantum entity is added to a collection of entities, its state space will be coupled, via the tensor product, to the state space of the collection, hence new collective states will emerge, primarily entangled ones, due to the superposition principle. These new entangled states are precisely those achieving the contextual updating in the quantum formalism, emphasizing the profound parallel between the quantum entanglement mechanism and that of meaning connections between conceptual entities.

\section{Time dilation\label{timedilation}}

To conclude our brief presentation of the power of the conceptuality interpretation to explain fundamental physical phenomena, we now also consider the theory of special relativity. Similar to quantum phenomena, relativity challenges classical assumptions, and here, too, the conceptuality interpretation is able to clarify the origin of relativistic effects.

A key result of relativity is that space, traditionally viewed as a static, encompassing realm, becomes problematic as a notion and is transformed into a personal and relational construct \citep{aerts2018,aertssassoli2024a}. Each physical entity, with its unique perspective, moves in a distinct space and more generally a distinct space-time. Thus, relativity, like quantum mechanics, suggests an underlying non-spatiotemporal realm from which the various individual spatiotemporal representations emerge. Therefore, the hypothesis that physical entities are fundamentally conceptual can also explain the nonspatiality inherent in the relativistic description.

To illustrate this, let's examine \emph{time dilation}, which, as we will show, arises naturally when time is viewed as being measured by cognitive entities counting the elementary steps in their thought processes. Think of two bodies, $A$ and $B$, not as objects but as conceptual entities engaging in meaning driven interactions. We also introduce two cognitive observers, $C_A$ and $C_B$, who focus their attention on the evolution of the meaning carried by $A$ and $B$, respectively. Assume for example that how the state of the conceptual entity $A$ changes reflects the way $C_A$ reflects on a problem, progressing from a \emph{Hypothesis} to a \emph{Conclusion}, through specific \emph{elementary conceptual steps}. The cognitive entity $C_B$, focusing on $B$'s evolution, follows a different cognitive path, and to place ourselves in the typical \emph{twin paradox} scenario, let us assume that $C_A$ and $C_B$ start with the same  \emph{Hypothesis} and reach the same \emph{Conclusion}. However, $C_A$ uses $n_A$ elementary conceptual steps to do so, while $C_B$ only uses $n_B < n_A$ steps, and let us assume for simplicity that $n_B = n_A/2$.

\begin{figure}
\begin{center}
\includegraphics[scale=0.22]{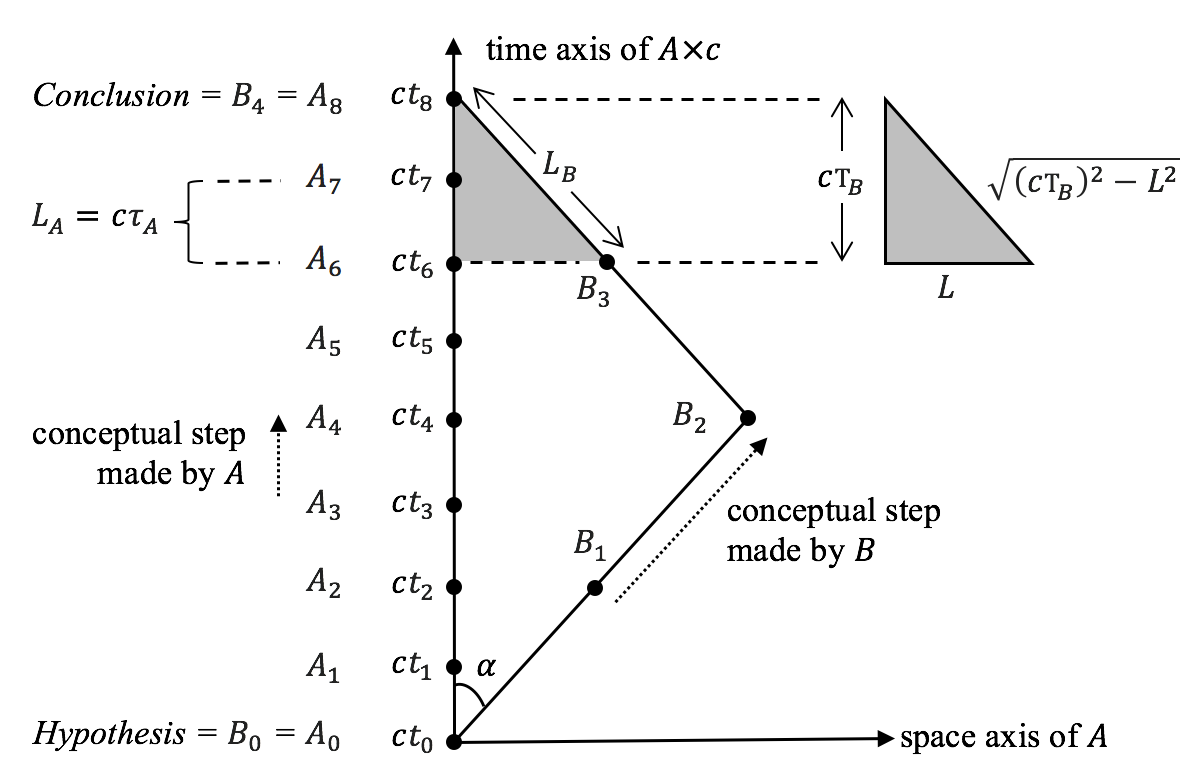}
\end{center}
\vspace{-15pt}
\caption{The coordination of the conceptual paths followed by the two entities $A$ and $B$, in the spacetime constructed by the cognitive observer $C_A$, here in the situation $n_A=8$ and $n_B=4$. \label{Figure3}}
\end{figure}

Suppose then that $C_A$ decides to represent each of $A$'s steps along an axis, assigning a unit length $L_A$ to each of them, and also a constant speed $c$, so the duration of each step is $\tau_A = L_A/c$. $C_A$'s reasoning then corresponds to a movement along this \emph{order parameter axis}, from a given initial point to a final one, situated at a distance $n_A L_A$ from the former. If $C_A$ decides to also focus on the evolution of $B$, since entities $A$ and $B$ are of the same nature, it will reasonably assume that they both produce cognitive steps at the same speed $c$. However, since the path followed by $B$, in the abstract conceptual realm, is such that it reaches the \emph{Conclusion} in half the steps used by $A$, $C_A$ cannot represent it on the same axis used to parameterize the path of $A$,  let us call it the \emph{time axis of $A$}, since the units on the latter have been chosen so that one needs twice as many steps to reach the \emph{Conclusion}.  To consistently parametrize also the evolution of $B$, $C_A$ is thus forced to introduce an additional axis,  let us call it the \emph{space axis of $A$}, and use this additionally introduced dimension to describe $B$ as moving on a outward and return path. Now, if we consider this construction from an \emph{Euclidean} perspective, things do not work. Indeed, the length of the $B$-path, calculated using the \emph{Pythagorean theorem}, will be longer than that of the $A$-path, but this would be inconsistent, since $B$ follows a shorter path, using only half the steps used by $A$; see Figure~\ref{Figure3}.

To solve this problem, $C_A$ must necessarily have recourse to a hyperbolic metric, e.g., the pseudo-Euclidean \emph{Minkowski metric}. It is then possible for the length $L_B$ of a single conceptual step of $B$ to be exactly equal to the length $L_A$ of a single conceptual step of $A$, i.e., to have the equality $L_A=L_B$, which is what $C_A$ wants, since the two entities $A$ and $B$ are assumed to change state at the same absolute speed $c$ in their common conceptual domain (see Figure~\ref{Figure3} and \citet{aerts2018,aertsetal2020}, for more details). In other words, by adopting a Minkowski metric, the cognitive observer $C_A$ is able to construct a personal space-time theater in which it can consistently track not only the cognitive process associated with $A$, but also that associated with $B$, and all it has to do is attach an appropriate spatial velocity $v$ to characterize the state changes of the latter. This means that the existence of the \emph{time dilation generalized parallax effects} becomes understandable and explainable when the existence of an underlying conceptual realm is taken into account. Hence, relativity, like quantum mechanics, suggests the existence of a conceptual realm in which entities all move at the same absolute speed $c$ \citep{aertssassoli2024a}.

\section{Conclusion\label{conclusion}}

In this second part of our two-part article, we have explored in a more systematic way some of the conceptuality interpretation's most clarifying explanations of quantum and relativistic behavior. The interpretation offers a compelling metaphysics in which entities do not intrinsically possess spatiotemporal attributes but acquire them emergently when large combinations of concepts are formed, corresponding to what we call \emph{stories} in our human culture. The latter would play a similar role as macroscopic objects play in the physical realm. This means entities formed by large collections of atoms would be similar to stories, but written in a non-human language.\footnote{This can be intuited by observing that if $A$ and $B$ are two concepts, the conjunction $A$ \emph{and} $B$ and the disjunction $A$ \emph{or} $B$ are also concepts. However, if $A$ and $B$ are combinations of numerous concepts, i.e., \emph{stories}, then although $A$ \emph{and} $B$ is still a story, formed by the concatenation of two stories, usually the disjunction $A$ \emph{or} $B$  is no longer considered to be a story, similarly to what happens to objects, as is clear that the conjunction of two objects is a composite object, but the disjunction of two objects is not an object, but a concept \citep{aertsetal2020}. This allows us to better understand the challenge of the so-called \emph{Heisenberg cut} in defining where the quantum realm ends and the classical realm begins. Translated into the human cultural realm, this is equivalent to asking when a text is large enough to become a story, which is, of course, a question that can only be answered contextually and not absolutely.} 

As we have seen, by embracing this cognitive-conceptual metaphysics, quantum mechanics becomes intuitive and intelligible, resolving many of the interpretive challenges that have persisted since the theory's inception. In this last section of our article, we briefly point out a more speculative idea suggested by the conceptuality interpretation, then conclude by situating it in relation to other interpretations. This idea is that there may be a duality in reality: on one side are the conceptual entities that carry meaning (with bosons as their archetypes), and on the other side are the cognitive entities that are sensitive to meaning (with structures made of fermions as their archetypes). And these two basic categories of entities would evolve symbiotically. The emerging worldview is thus a \emph{pancognitivist} one, where everything in reality would participate in cognition, with human cognition being just one example of it, expressed at a specific organizational level. If this is true, then, quoting from \citet{aertssassoli2018}:

{\quote  [...] something similar to what happened in our human macro-world, with individuals using concepts and their combinations to communicate, may have already occurred, and continue to occur, \emph{mutatis mutandis}, in the micro-realm, with the entities made of ordinary matter communicating and co-evolving thanks to a communication that uses a language made of concepts and combinations of concepts that are precisely the quantum entities and their combinations.\endquote}

This remains, for now, a speculative perspective that calls for further critical investigation. Yet, it is a compelling view with significant implications for our understanding of reality and evolution as a whole. Indeed, if the interpretation proves correct, the framework needed to describe the evolving physical reality would be one of  \emph{cultural evolution}. This implies that what we consider on our planet to be a secondary evolutionary process, emerging after the evolution of biological species, would actually be a far more ancient process of change and, in a sense, the primary process of transformation that has governed our universe from its inception \citep{aertssassoli2018,aertssassoli2022}.

That being said, it is natural to ask how the conceptuality interpretation, with its pancognitivist view, relates to panpsychism \citep{pan2022,gao2013}, panintentionalism \citep{barros2022}, and to the more general debate about consciousness, in particular the ``consciousness causes collapse'' interpretations \citep{chalmers2013,stapp2009,gornitz2018}. In short, the conceptuality interpretation is compatible with, but slightly different from, panpsychism in that it does not state, as a top-down principle, that mind or consciousness would be a fundamental feature of the world. Its position is more pragmatic and bottom-up. It observes, in its quest to understand and explain the phenomena revealed by quantum mechanics and relativity, that cognitive processes are much more prevalent in physical reality than we initially imagined, contributing to its organization not only in the human cultural realm, but also in the realm of matter and energy. Thus, it posits that our best explanation of the observed behaviors is to consider that the elementary components of the universe are conceptual in nature. This means that physical processes would be governed by meaning, but this does not require consciousness per se. It is, in principle, possible for cognition to exist without the presence of consciousness. While this idea may seem far-fetched, cognition and the conceptuality underlying it could, in theory, emerge through natural selection, as they represent the most efficient mode of interaction. In such a scenario, consciousness could be both unnecessary and absent. Thus, for the time being, the conceptuality interpretation remains closer to the more parsimonious view of panintentionalism \citep{barros2022}, since it does not require the qualitative character of conscious experiences. However, we wish to emphasize that we do not rule out the possibility that consciousness may ultimately play a significant role in the cognitive processes we have suggested take place in the quantum realm. In the near future, we aim to provide a more concrete articulation of this potential role, which we believe is plausible given certain technical aspects of our formalism. As with our interpretation itself, this future exploration of consciousness within the conceptuality interpretation will also follow a bottom-up approach, grounded as much as possible in the available evidence.

Let us conclude by asking how the interpretation stands in relation to the other available interpretations. Fully answering this question would require an entire article, so here we will simply observe how the interpretation positions itself with respect to the three claims proposed by Maudlin \citep{maudlin1995}, which are often used to categorize the different 
interpretations. The first claim is that of the completeness of the wave function of a system, i.e. the assumption that it would directly or indirectly specify all properties of a system and thus its complete state. For the conceptuality interpretation this claim is partly valid, in the sense that in many situations a Hilbert space vector correctly describes the \emph{state of affairs} of a system by specifying all its properties, both actual and potential. However, such a specification is not always sufficient, for at least two reasons. The first is that density operators would also play a role in describing the genuine state of a system, either during a measurement process or, more generally, when a system is entangled with another system \citep{AertsSassoli2016}. The second reason is that the interpretation does not necessarily ascribe a fundamental role to the Hilbert space geometry, since one can observe in the human cognitive domain violations of some of the relations resulting from it and the Born rule, such as the so-called marginal laws \citep{aabgssv2019}. This means that the Hilbert space linear structure can only be regarded as an approximation of a more general formalism, in which, for example, separation can also be described \citep{aertssassolisozzoproceedings2024}. Maudlin's second statement corresponds to the hypothesis that the wave function would evolve only deterministically, e.g., by the unitary evolution derived from the Schr\"{o}dinger equation. The conceptuality interpretation does not consider this statement to be valid. In fact, in a pancognitivist reality, cognitive processes, and especially decision-making processes, are ubiquitous, so that evolution is mainly nonlinear and characterized by a wide range of possible outcomes, at any given time. In this sense, unitary evolution would only be a special case of a more general evolution law, corresponding to the presence of only one possible outcome. And with this observation, we arrive at Maudlin's third statement, which expresses the existence of truly indeterministic measurement contexts, characterized by multiple possible outcomes, only one of which is actualized at any given time, according to the relative frequencies dictated by the Born rule. The conceptuality interpretation agrees with this, being one-world interpretation in which only a single outcome is actualized at a time. However, as already discussed, Born's rule is seen as a first-order approximation of a more general probabilistic rule \citep{AertsSassoli2015a,AertsSassoli2014}.

\section*{Acknowledgements}
This work was supported by the project ``New Methodologies for Information Access and Retrieval with Applications to the Digital Humanities'', scientist in charge S. Sozzo, financed within the fund ``DIUM -- Department of Excellence 2023--27'' and by the funds that remained at the Vrije Universiteit Brussel at the completion of the ``QUARTZ (Quantum Information Access and Retrieval Theory)'' project, part of the ``Marie Sklodowska-Curie Innovative Training Network 721321'' of the ``European Unions Horizon 2020'' research and innovation program, with Diederik Aerts as principle investigator for the Brussels part of the network.

\end{document}